\documentclass[%
 aip,
 sd,%
 amsmath,amssymb,
 reprint,%
]{revtex4-1}

\usepackage{graphicx}  
\usepackage{dcolumn}   
\usepackage{bm}        
\usepackage{graphicx}
\usepackage{subfig}
\usepackage{epsfig}
\usepackage{epstopdf}
\usepackage{enumerate}
\usepackage{physics}
\newcommand{\RNum}[1]{\uppercase\expandafter{\romannumeral #1\relax}}

\usepackage{url}

\begin{document}

\preprint{AIP/123-QED}

\title{Role of Coherence in Excitation Transfer Efficiency to the reaction center in Photosynthetic Bacteria {\it Chlorobium Tepidum}}

\author{Davinder Singh}
\email{davinder.singh@iitrpr.ac.in}

\author{Shubhrangshu Dasgupta}%
 
\affiliation{Department of Physics, Indian Institute of Technology Ropar, Rupnagar, Punjab - 140001, India}

\date{\today}   


\begin{abstract}
We investigate the effect of coherence to the excitation transfer efficiency (ETE) in photosynthetic bacteria {\it Chlorobium Tepidum}. We have modeled the monomer of Fenna-Matthews-Olson (FMO) complex as consisting of eight bacteriochlorophyll-a sites, while explicitly consider reaction center core complex (RCC) as an additional site. With the use of realistic bath spectrum and several dominant vibronic modes in the non-Markovian master equation, in an effective 9-site model, we have compared the ETE for an initial pure state and an initial mixed state. We observe that the initial pure state relaxes efficiently to increase the trapping at the RCC. We further illustrate that the coherence play a competitive role to block the back transfer of excitation from RCC pigment to FMO complex and hence to maximize the ETE.
\end{abstract}

\pacs{82.20Wt, 82.20Rp, 82.53Ps, 87.15H-}    
\keywords{Excitation energy transfer, Coherence}    
\maketitle


\section{Introduction}

During the light-harvesting by photosynthetic green sulfur bacteria {\it Chlorobium Tepidum}, a solar photon is absorbed by pigments of chlorosome antenna. Subsequently, this excitation energy is transferred with very high quantum efficiency ($\sim 100 \%$) to the reaction center (where the excitation energy is trapped\cite{hauska_reaction_2001}) via Fenna-Matthews-Olson (FMO) complex\cite{schmidt_am_busch_eighth_2011}. At the reaction center core complex (RCC), the separation of electron leads to the highly efficient trapping of excitation\cite{Amerongen,Blankenship}. The mechanism that blocks the transfer of excitation from the RCC to the FMO complex backwards is however still not known and so is the reason for the large quantum efficiency, which could be used to improve the solar light-harvesting to meet our future renewable energy needs \cite{Meyer_Chemist_2011,Gust_Solar_2009,Maeda_Photocatalytic_2010,Mallouk_The_2010,
Vullev_From_2011,Gust_Mimicking_2010}.

Recently the excitation energy transfer in FMO complex has been studied using 2-D electronic spectroscopy\cite{Brixner} (2DES)\cite{Engel_evidence_2007,Panitchayangkoon_long_2010,panitchayangkoon_direct_2011}. It has been observed that the transfer of excitation is an oscillatory coherent process rather than an incoherent one as had been predicted by earlier theoretical models \cite{lambert_quantum_2013,scholes_lessons_2011,collini_coherently_2010,scholes_quantum-coherent_2010}. 

Following these observations, the effect of quantum coherence on excitation transfer efficiency (ETE) has been studied by a number of theoretical approaches. Using Markovian master equation approach, Aspuru Guzik {\it et. al.} have studied the excitation trapping at lowest energy site of FMO complex \cite{rebentrost_environment-assisted_2009,mohseni_environment-assisted_2008,rebentrost_role_2009}. Further, Plenio {\it et. al.} have modelled the excitation transfer as dissipation of connected complex network to the implicitly connected reaction center \cite{plenio_dephasing-assisted_2008,caruso_highly_2009,chin_noise-assisted_2010}. The effect of environmental de-phasing\cite{leon-montiel_highly_2013} and randomly selected initial excitations\cite{shabani_numerical_2014} has also been studied for ETE. It has been reported that the ETE strongly depends on the coupling to the protein environment \cite{rebentrost_environment-assisted_2009,mohseni_environment-assisted_2008,rebentrost_role_2009,plenio_dephasing-assisted_2008,caruso_highly_2009,chin_noise-assisted_2010,shabani_numerical_2014}. It is further shown that the enhancement of ETE by quantum coherence is negligible \cite{wilkins_why_2015,briggs_equivalence_2011}. However, interestingly the above theoretical studies have not explicitly considered the RCC while studying ETE. In this paper, we treat the RCC as an additional site and explore the effect of coherence in trapping excitation at RCC, and thereby in ETE. We consider several other important features, as follows, in a consolidated way, one or more of which are missed in earlier theoretical studies:

\begin{enumerate}[(i)]

\item The previous studies, as mentioned above, have used only 7-site model of the monomer of FMO complex, while each monomer is known to consist of 8 bacteriochlorophyll-a (BChla) sites \cite{ben-shem_evolution_2004,tronrud_structural_2009}. We choose all these sites and their suitable internal couplings. 
\item It is shown that the dynamical coherences originate from vibronic modes local to each BChla site \cite{chenu_enhancement_2013,kreisbeck_disentangling_2013}. Suitable set of vibronic modes with varied coupling constants have been considered in the present study. 
\item Further, in ensemble measurements, the initial condition plays a very important role\cite{cao_correlations_2006,eisfeld_phase_2011}. We strongly emphasize that to systematically study the effect of coherence on ETE, the effect of initial coherence must be included in the dynamics. 
\item The X-ray crystallographic studies of FMO complex have reported that the BChla sites of FMO complex are surrounded by inhomogeneous distribution of protein molecules \cite{tronrud_structural_2009,milder_revisiting_2010}. However, to study the ETE, earlier theoretical models have considered only a homogeneous protein environment around all the BChla sites to conclude that the environment assists the transport of excitation energy in FMO complex \cite{rebentrost_environment-assisted_2009,mohseni_environment-assisted_2008,rebentrost_role_2009,plenio_dephasing-assisted_2008,caruso_highly_2009,chin_noise-assisted_2010,leon-montiel_highly_2013,shabani_numerical_2014}. In this respect, inhomogeneity of protein environment must be included in the dynamics to study its effect more realistically.

\end{enumerate}

The structure of the paper is as follows: In Section \RNum{2}, we describe the model to study ETE in FMO-RCC. In the following section (Section \RNum{3}), we present the results of our numerical study followed by a detailed discussion. Finally in Section \RNum{4} we conclude with summary and an outlook.


\section{Model}
\begin{figure}[!h]
\begin{center}
\subfloat[]{\label{In_situ_Interaction}
\includegraphics[width = 2 in]{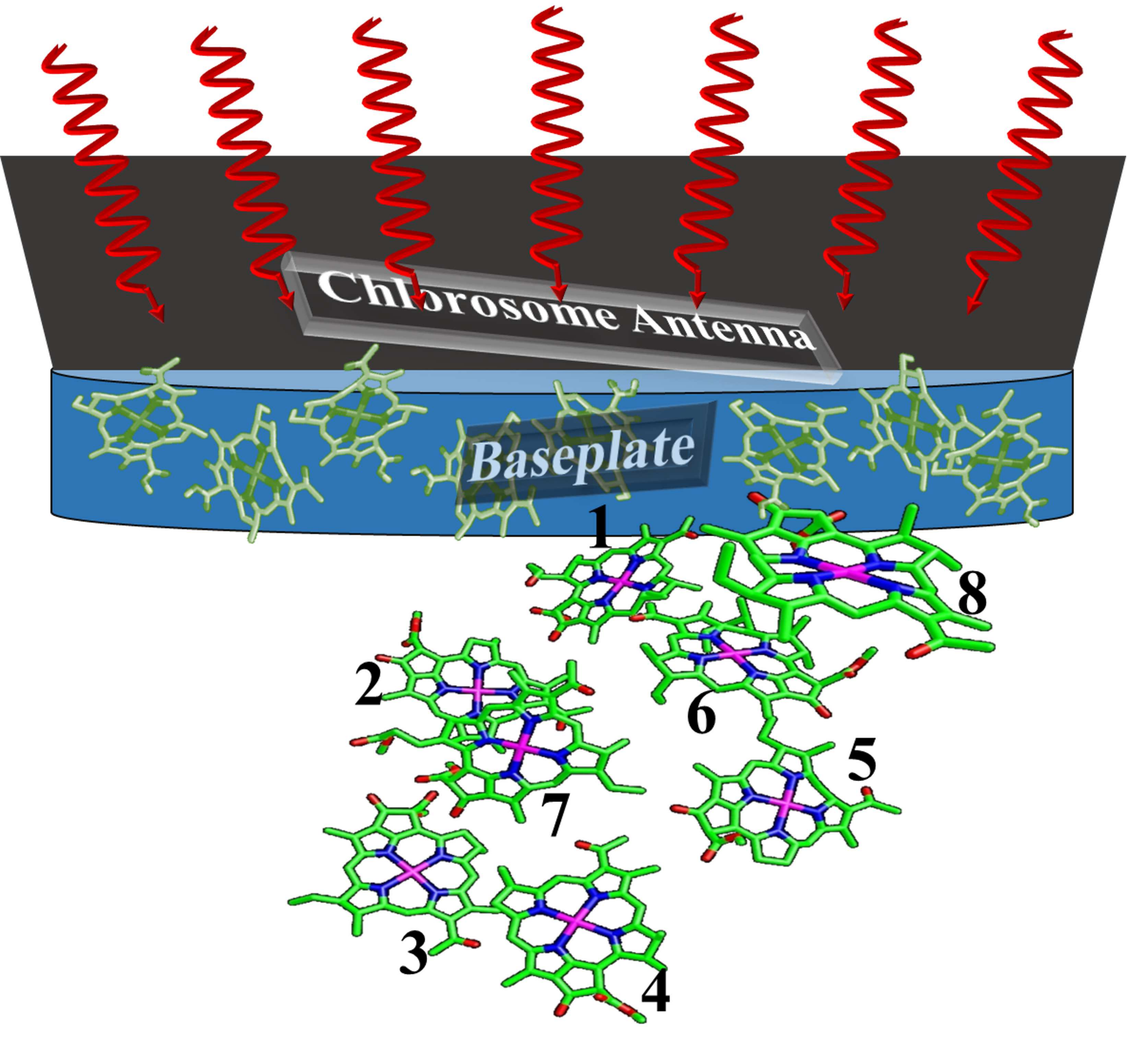}}
\quad
\subfloat[]{\label{Laser_Interaction}
\includegraphics[width = 2 in]{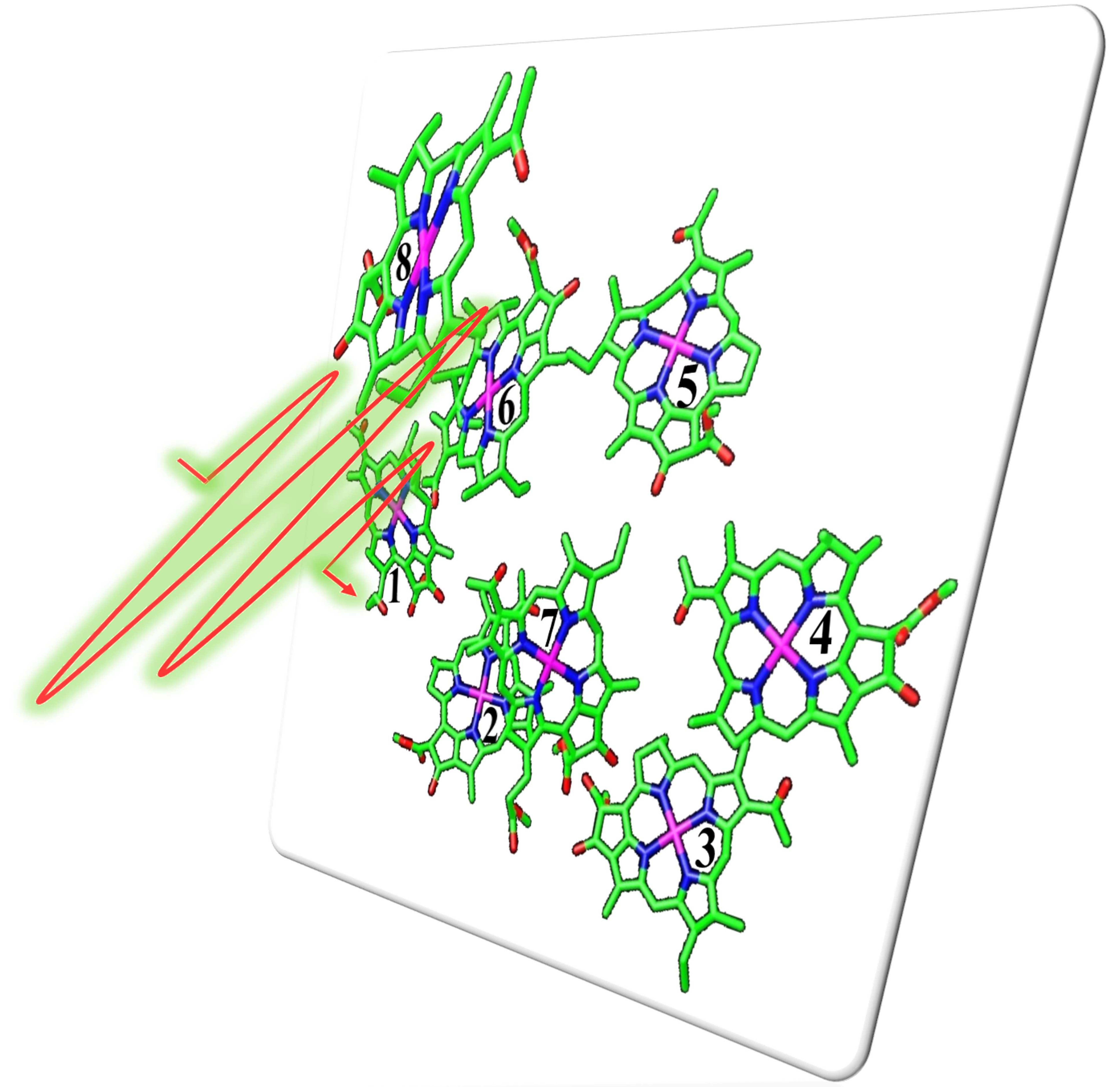}}
\quad
\subfloat[]{\label{Levels_nine}
\includegraphics[width = 2 in]{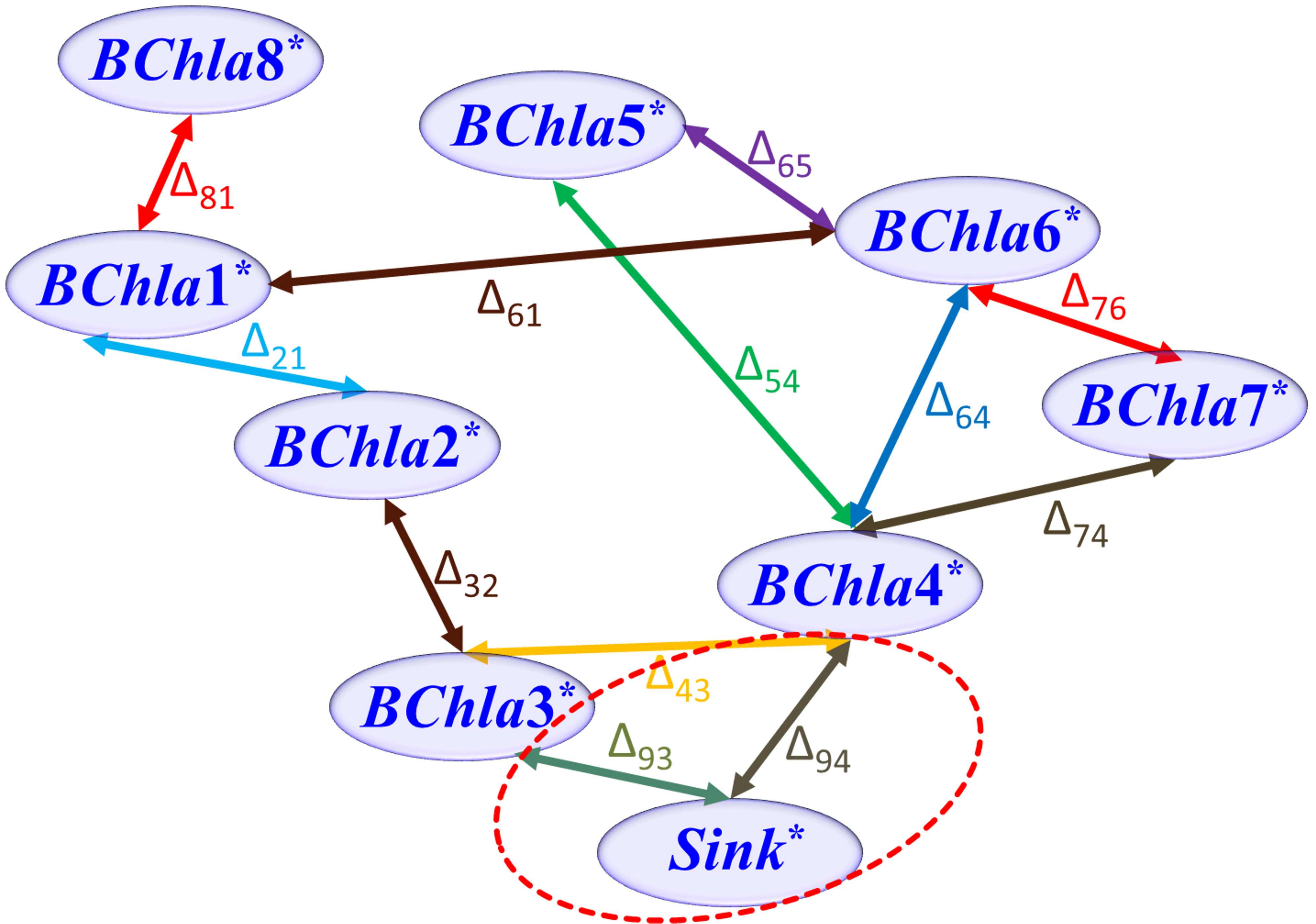}}
\caption{(Color online)(a) Schematic representation of incoherent {\it in-vivo} absorption of light by chlorosome antenna followed by transfer of excitation to the monomer of FMO complex via base-plate BChla sites. (b) Schematic illustration of interaction of femto-second laser pulse with monomer of {\it isolated} FMO complex. The figures were created using VEGA-ZZ for the PDB entry 3EOJ. (c) Schematic illustration of active channels of transfer of excitation between excited pigments.}
\label{Absorption}
\end{center}
\end{figure}

In FMO complex, the excitation transfer is almost negligible between two adjacent monomers\cite{ritschel_absence_2011}, and it occurs predominantly within one sub-unit only. To study the excitation transfer dynamics, each pigment is modelled as a two-level system. In pigment-protein complexes, the formation of bi-excitons and other higher order excitons is blocked to avoid the dissipation of excitation energy\cite{dong_photon_2017}. Following this, we have used only singly excited subspace and the active channels of excitation transfer are illustrated in Fig.\ref{Levels_nine}. To study the ETE explicitly, we have also considered an excited pigment of RCC as the ninth site as well as a sink ( see Fig.\ref{Levels_nine}). Here we assume that sink is spatially close to the BChla 3 and BChla 4 with the highest spectral overlap and this FMO complex is energetically coupled to RCC.

In the singly excited subspace, the system Hamiltonian can be written as\cite{May}

\begin{equation}
H_{S} = \sum_{j = 1}^{9} \hbar \epsilon_{j}\ket{j}\bra{j} + \sum_{i<j}\Delta_{ji}\left( \ket{j}\bra{i} + \ket{i}\bra{j}\right) \;,
\label{System_Hamiltonian}
\end{equation}
where $\ket{j}$ represents the excited state of j$th$ site and $\epsilon_{j}$ is the difference of transition frequency of j$th$ BChla site with respect to the BChla site with the lowest transition frequency. The tunneling frequency between j$th$ and i$th$ BChla sites is illustrated by $\Delta_{ji}$. Note that, the site $\ket{9}$ (the sink) represents the pigment of RCC.

Further, $H_{B}$ denotes the bath Hamiltonian
\begin{equation}
H_{B} = \sum\limits_{k_{j}} \hbar \omega_{k_{j}}b_{k_{j}}^\dagger b_{k_{j}}\;,
\label{Bath_Hamiltonian}
\end{equation}
where, $b_{k_{j}}$ and $b_{k_{j}}^\dagger$ are the annihilation and the creation operators, respectively, for the $k_{j}$th bath mode, local to $j$th BChla site.

Finally, the Hamiltonian of system-bath interaction can be described as
\begin{equation}
H_{SB} = \frac{\hbar}{2}\sum_{j} \sum\limits_{k_{j}} g_{k_{j}}\ket{j}\bra{j}(b_{k_{j}} + b_{k_{j}}^\dagger) \;,
\label{Interaction_Hamiltonian}
\end{equation}
where $g_{k_{j}}$ is the coupling of environmental phonons local to $j$th BChla site. 

Besides the reversible coherent evolution of excitation energy, two irreversible processes also occur which result in the loss of excitation from the FMO complex. One such process is the exciton recombination, responsible the dissipation of excitation to protein environment. The exciton recombination can be described by the following Hamiltonian \cite{rebentrost_role_2009}

\begin{equation}
-iH_{dissipation} = -i\hbar \Gamma\sum_{j}\ket{j}\bra{j}\;,
\label{dissipation_Hamiltonian}
\end{equation}
where $\Gamma$ represents the inverse life time of exciton.

The other irreversible process describes the trapping of excitation at RCC associated with the charge separation. The corresponding trapping Hamiltonian can be written as \cite{rebentrost_role_2009}
\begin{equation}
-iH_{trapping} = -i\hbar \Lambda_{9}\ket{9}\bra{9}\;,
\label{trapping_Hamiltonian}
\end{equation}
where $\Lambda_{9}$ is the trapping rate at the pigment of RCC.

Considering the above loss processes, the final non-hermitian Hamiltonian for ETE can be written as

\begin{equation}
H = H_{S} + H_{B} + H_{SB} - iH_{dissipation} - iH_{trapping}\;.
\label{Hamiltonian}
\end{equation}

We choose the density matrix approach to study the dynamics of excitation transfer. The dynamics of the reduced density matrix of the system can be described by the following non-Markovian master equation \cite{singh_coherence_2017}:
\begin{equation}
\begin{array}{lll}
\dot{\rho} &=& -\frac{i}{\hbar}\left[ H_{S},\rho\right]\\
&+&\sum_{j}\left\{\left( \ket{j}\rho_{jj}\bra{j} - \rho\ket{j}\bra{j}\right)D_{j}(t)\right. \\
&+&\left( \ket{j}\rho_{jj}\bra{j} - \ket{j}\bra{j}\rho\right)D_{j}^{\ast}(t)\\
&+&\left( \ket{j}\rho_{jj}\bra{j} - \ket{j}\bra{j}\rho\right)U_{j}(t)\\
&+&\left( \ket{j}\rho_{jj}\bra{j} - \rho\ket{j}\bra{j}\right)U_{j}^{\ast}(t)\\
&-&\left(\ket{j}\bra{j}\rho + \rho\ket{j}\bra{j}\right)\Gamma \\
&-&\left. \left(\ket{9}\bra{9}\rho + \rho\ket{9}\bra{9}\right)\Lambda_{9} \right\} \;.
\end{array}
\label{non-Markovian}
\end{equation}
The time-dependent coefficients in Eq.(\ref{non-Markovian}) describe the system-bath correlations as follows
\begin{eqnarray}
\label{correlation1}D_{j}(t) &=& \int_{0}^{t}dt^{\prime}\int_{0}^{\infty} d\omega J_{j}(\omega)\bar{n}(\omega,T)e^{-i\omega(t-t^{\prime})}\;,\\
\label{correlation2}U_{j}(t) &=& \int_{0}^{t}dt^{\prime}\int_{0}^{\infty} d\omega J_{j}(\omega)[\bar{n}(\omega,T) + 1]e^{-i\omega(t-t^{\prime})}\;.
\end{eqnarray}

Here $\bar{n}(\omega,T)$ is the average number of phonons of frequency $\omega$ at temperature T. Note that in the master equation approach, the information about the vibrational modes of environmental phonos and vibronic modes of BChla sites can be specified by the spectral density function $J_{j}(\omega)$ \cite{May}. The spectral density function used in this study has been obtained by the fitting with fluorescence line narrowing spectroscopy data\cite{wendling_electronvibrational_2000} and can be written as follows
\begin{equation}
J_{j}(\omega) = K_{j}\omega \left( \frac{\omega}{\omega_{c_{j}}}\right)^{-1/2} e^{-\frac{\omega}{\omega_{c_{j}}}} + \sum_l K_{l} e^{-\frac{(\omega - \omega_{l})^{2}}{2d^{2}}}\;.
\label{Spectral-Density}
\end{equation}
Here $K_{j}$ (with $K_{j} = g_{k_{j}}^{2}$) represents the Huang-Rhys factor of vibrational modes of environmental phonons and $K_{l}$ describes the Huang-Rhys factor of vibronic modes local to each BChla site. Further the cutoff frequency of environmental phonons is illustrated by $\omega_{c_{j}}$. The frequency of vibronic modes is represented by $\omega_l$. In our analysis, the following vibronic modes with non-negligible values of Franck-Condon factors have been used: $\omega_{l}$ = 36 cm$^{-1}$, 70 cm$^{-1}$, 173 cm$^{-1}$, 185 cm$^{-1}$, and 195 cm$^{-1}$. We have choosen the values of $K_{l}$ to be 40 times to that of the corresponding Franck-Condon factors given by Wendling {\it et al.} \cite{wendling_electronvibrational_2000}, and the width of the vibronic band as $d = \sqrt{18}\omega_{0}$ (where $\omega_{0} = 100$ cm$^{-1}$, is used for the normalization). From the fitting with the experimental data, the observed values of $K_{j}$ and $\omega_{c_{j}}$ are listed in Table\ref{Table_EET}. Different values of $K_{j}$ and $\omega_{c_{j}}$ for different BChla sites describe the in-homogeneity of protein environment as has been observed with the relevant X-ray crystallography experiments.

\begin{table}[h!]
\caption{\label{Table_EET}Parameters used to simulate the ETE in FMO-RCC complex.}
\begin{ruledtabular}
\begin{tabular}{lr}
\textrm{Electron-phonon couplings}&
\textrm{Cut-off frequencies (in cm$^{-1}$)}\\
\colrule
$K_{1}$ = 0.0510 & $\omega_{c_{1}}$ = $45.6577 $ \\ 
$K_{2}$ = 0.0425 & $\omega_{c_{2}}$ = $54.7823 $ \\ 
$K_{3}$ = 0.0860 & $\omega_{c_{3}}$ = $27.1026 $ \\ 
$K_{4}$ = 0.0600 & $\omega_{c_{4}}$ = $38.5205 $ \\  
$K_{5}$ = 0.0486 & $\omega_{c_{5}}$ = $47.9760 $ \\ 
$K_{6}$ = 0.0460 & $\omega_{c_{6}}$ = $50.6508 $ \\ 
$K_{7}$ = 0.0525 & $\omega_{c_{7}}$ = $44.3918 $ \\ 
$K_{8}$ = 0.0365 & $\omega_{c_{8}}$ = $63.8000 $ \\ 
$K_{9}$ = 0.0566 & $\omega_{c_{9}}$ = $41.1554 $ \\ 

\end{tabular}
\end{ruledtabular}
\end{table}

Next, we numerically solve Eq.(\ref{non-Markovian}) for different initial conditions and for the occupation probabilities ("population") in different sites. It is important here to note that {\it in-situ} absorption of light by photosynthetic bacteria is generally different from the absorption of coherent laser light in an {\it isolated} FMO complex. In the earlier case, incoherent photons are absorbed by the chlorosome antenna pigments ( see Fig.\ref{In_situ_Interaction}) and this energy is then transferred to FMO complex which results in a mixed state \cite{schmidt_am_busch_eighth_2011}. For the {\it in-vivo} incoherent absorption, we use the following mixed state as the initial condition\cite{schmidt_am_busch_eighth_2011}:
\begin{equation*}
\rho_{11} = 0.29; \rho_{22} = 0.09; \rho_{33} = 0 ; \rho_{44} = .01;
\end{equation*}
\begin{equation*}
\rho_{55} = .026; \rho_{66} = .017; \rho_{77} = .017; \rho_{88} = .55 
\end{equation*}
\begin{equation}
\rho_{ij} = 0; \qquad i \neq j\;.
\end{equation}

On the other hand, in the later case, a coherent laser pulse interacts with an {\it isolated} FMO complex (Fig.\ref{Laser_Interaction}) and creates a coherent pure state. We have used the following initial coherent pure state also as described in detail by the authors\cite{singh_effect_2017}
\begin{equation}
\begin{array}{lll}
\ket{\psi} &=& \left. \sqrt{0.3260}\ket{1} + \sqrt{0.0795}\ket{2} + \sqrt{0.0005}\ket{3} +\sqrt{0.2641}\ket{4} \right.\\
&+& \left. \sqrt{0.0321}\ket{5} + \sqrt{0.0127}\ket{6} + \sqrt{0.2826}\ket{7} + \sqrt{0.0025}\ket{8} \right.\;.
\end{array}
\label{Pure_State}
\end{equation}

To understand the dynamics of ETE and the trapping of excitation at RCC, we quantify the ETE as the integrated probability of excitation entrapment by RCC \cite{rebentrost_role_2009} as follows

\begin{equation}
\eta(t) = 2\Lambda_{9}\int_{0}^{t}\rho_{99}(t^{\prime})dt^{\prime}\;.
\label{Efficiency}
\end{equation}

\section{Results and discussions}
\subsection{Effect of coherence}
\begin{figure}[!h]
\begin{center}
\subfloat[]{\label{Sink_277_3_enrgy_sink}
\includegraphics[width = 3.0 in]{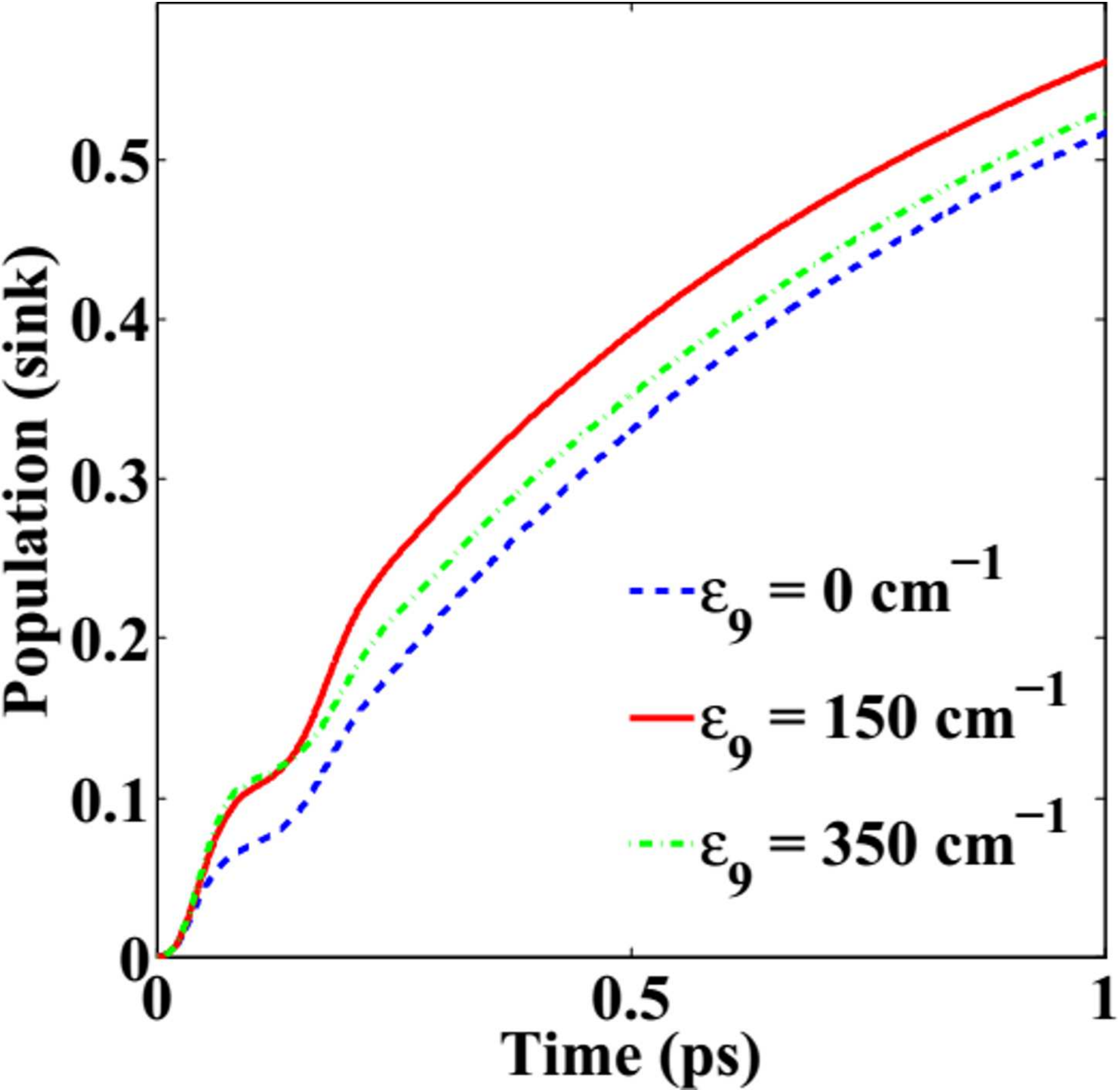}}
\quad
\subfloat[]{\label{Sink_277_3_coherent_cohrnce_comparison_final}
\includegraphics[width = 2.9 in]{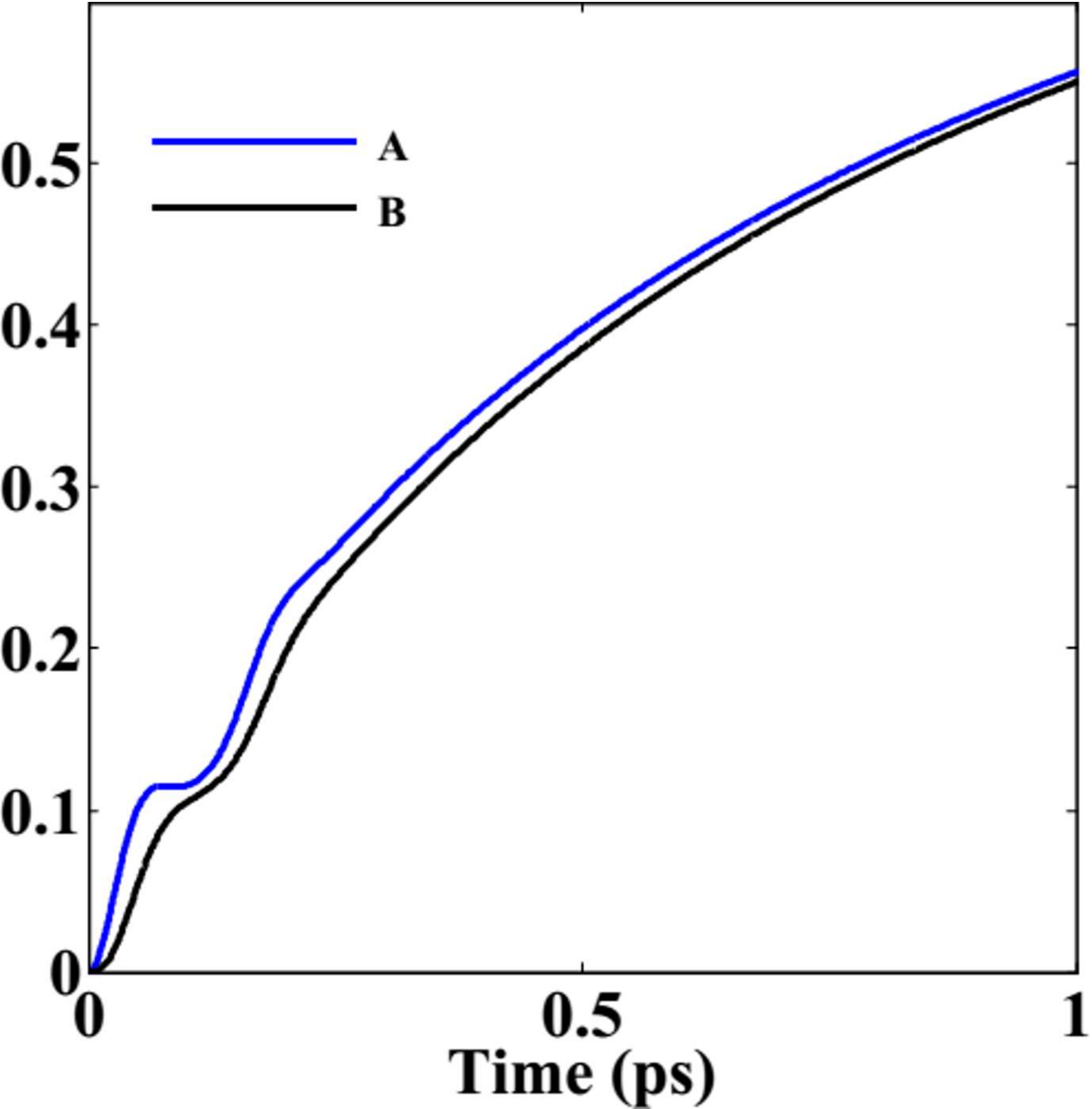}}
\caption{(Color online) Dynamics of sink population ($\dot{\eta} = 2 \Lambda_{9}\rho_{99}$) at room temperature T = 277 K in FMO-RCC complex. (a) Different values of $\epsilon_{9}$ have been considered to study the time evolution. (b) Illustration of the separate contributions of coherence terms only (i.e $A$) and population terms only (i.e. $B$) in the population dynamics of $\rho_{99}$. We have used $\Lambda_{9} = 3$ ps$^{-1}$ and $\Gamma = 1$ ns$^{-1}$.}
\end{center}
\end{figure}

For the FMO complex, we use the values of transition frequencies ($\epsilon_{j}$) and tunneling frequencies ($\Delta_{ji}$) as reported by Renger et al.\cite{schmidt_am_busch_eighth_2011}. Note that, the relative value of transition frequency of pigments of RCC (i.e $\epsilon_{9}$) as compared to the transition frequencies of BChla sites are not known. Here, we have chosen different values of $\epsilon_{9}$ and simulate the dynamics for ETE as shown by Fig.\ref{Sink_277_3_enrgy_sink}. Our dynamical simulation shows that when we choose $\epsilon_{9} = 0$ cm$^{-1}$ (condition favorable for energetic trap), the ETE is less as compared to the case when $\epsilon_{9}$ is larger ($\equiv 150$ cm$^{-1}$) [see Fig.\ref{Sink_277_3_enrgy_sink}]. Moreover, a further increase of $\epsilon_{9}$ does not lead to larger ETE. Clearly, there is an optimum value of $\epsilon_{9}$ that leads to maximum ETE. It further implies that it is not the energetic trap at sink as had been assumed by the earlier theoretical model\cite{Ishizaki}, rather the quantum interference between two channels of transfer of excitation to the sink from BChla 3 and BChla 4 as shown in Fig.\ref{Levels_nine} helps to trap the excitation at sink efficiently. 

To further verify the explicit role of interference on the excitation trapping at sink, we have studied the evolution of $\rho_{99}$, which is governed by the following equation
\begin{equation}
\dot{\rho}_{99} = - 2\Delta_{93}\rho_{93}^{I} - 2\Delta_{94}\rho_{94}^{I} - 2(\Gamma + \Lambda_{9})\rho_{99}\;,
\label{sink_population_eq}
\end{equation} 
where $\rho_{93}^{I}$ and $\rho_{94}^{I}$ represent the imaginary parts of the respective coherences. This can also be written as
\begin{equation}
\dot{\rho}_{99} =  A - B\;,
\label{sink_population_eq1}
\end{equation} 

\noindent where $A$ is the contribution from coherence terms only and $B$ contains the contribution of the population $\rho_{99}$. Note that in absence of the term A, this would provide an exponentially decaying solution to $\rho_{99}(t)$. However, as shown in Fig.\ref{Sink_277_3_coherent_cohrnce_comparison_final}, it turns out that A remains greater than B during the evolution, making $\dot{\rho}_{99} > 0$. This indicates that the population in the 9th site, i.e., the sink, is ever increasing and there is no back flow of excitation from the sink (which would correspond to $\dot{\rho}_{99} < 0$). This is clearly an effect of the term A, which contain the coherence terms only.  Precisely speaking, as imaginary parts of $\rho_{93}$ ( i.e. $\rho_{93}^{I}$) and $\rho_{94}$ ( i.e. $\rho_{94}^{I}$) become negative (because $A$ remains positive), this refers to a one-way excitation transfer of population in the channels 3$\rightarrow$9 and 4$\rightarrow$9 at a higher rate than the decay from the level 9. The dominance of the coherence terms suggests that the back transfer of excitation from the sink to the BChla 3 and BChla 4 is blocked due to interference between these two channels highlighted in Fig.\ref{Levels_nine}. Such a blocking essentially leads to enhancement in ETE. In the rest of the numerical simulation, we choose $\epsilon_{9} = 150$ cm$^{-1}$, corresponding to maximum ETE.
 
\subsection{Effect of initial condition}

\begin{figure}[!h]
\begin{center}
\subfloat[]{\label{Sink_277_3_Incoherent_Long_new}
\includegraphics[width = 2.7 in]{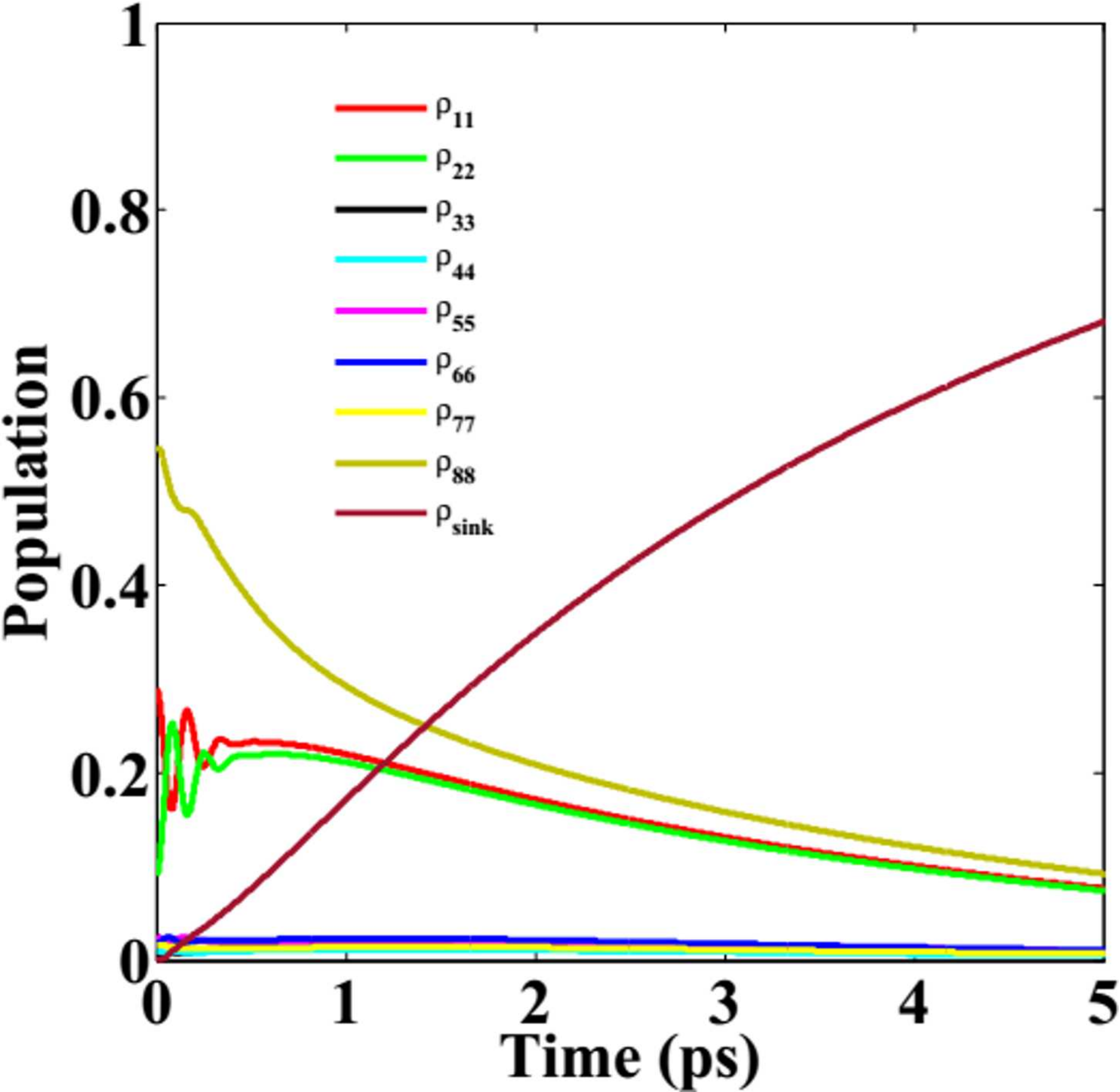}}
\quad
\subfloat[]{\label{sink_277_3_Coherent_long_new}
\includegraphics[width = 2.7 in]{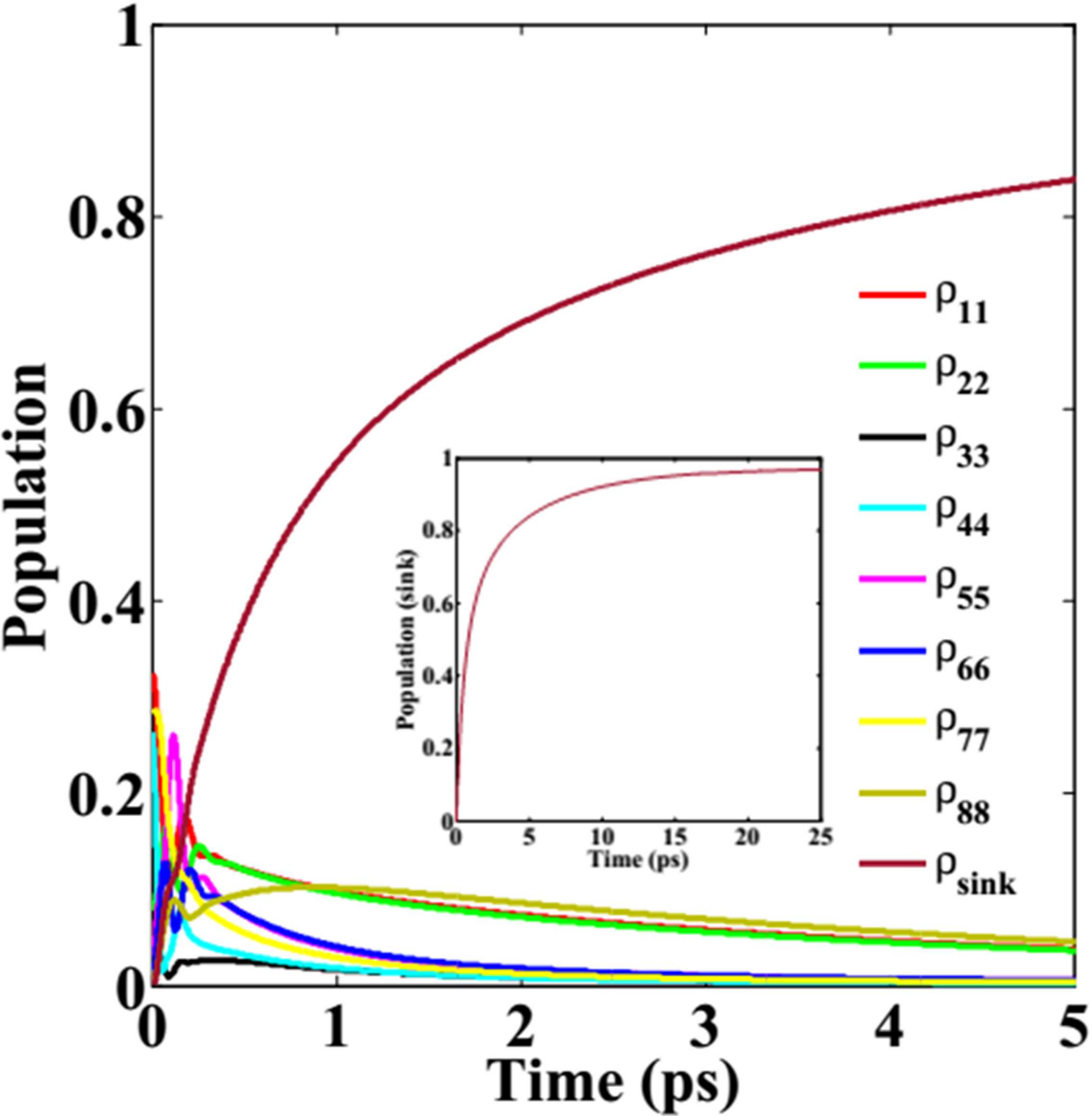}}
\caption{(Color online) (a) Dynamics of site population with an initial mixed state and (b) with an initial pure state. (Inset) Long time evolution with an initial pure state.}
\label{Sink_277_3_Long}
\end{center}
\end{figure}

We next analyze the effect of initial condition on the ETE. The long-time simulation shows that for incoherent initial mixed state the ETE is 64$\%$ at a time scale of 5 ps, typical time of transfer of excitation in FMO complex\cite{chin_noise-assisted_2010}, as shown in Fig.\ref{Sink_277_3_Incoherent_Long_new}. But for an initial pure state, the ETE is 84$\%$ at the same time scale (Fig.\ref{sink_277_3_Coherent_long_new}). It implies that the initial coherence enhances the ETE by almost 20$\%$. At the steady state, for the initial coherent pure state the ETE is 97$\%$ as shown by Fig.\ref{sink_277_3_Coherent_long_new}(inset), which is consistent with the observation of almost unity quantum efficiency\cite{schmidt_am_busch_eighth_2011,chin_noise-assisted_2010}. 

The enhancement of ETE for the initial coherent pure state indicates toward the possibility of two situations: (a) the preservation of memory by non-Markovianity enhances the ETE or (b) the initial pure state relaxes efficiently to increase the trapping at RCC. To demonstrate the effect of non-Markovianity, we have compared the ETE with both the Markovian and non-Markovian master equation. In the Markovian master equation the time dependent co-efficients of Eq.(\ref{correlation1}) and (\ref{correlation2}) become time-independent de-phasing rate $\gamma$. Our dynamical simulation shows that even for Markovian dynamics the ETE is almost the same (i.e 81 $\%$) for the initial pure state as it happens for non-Markovian case. It implies that the non-Markovianity has marginal effect on the additional enhancement of ETE. Rather it is the initial coherent pure state which relaxes efficiently to increase the trapping at sink. This efficient relaxation of initial pure state occurs because the interaction of femtosecond laser pulse, to prepare the initial pure state, coherently couples the excited BChla sites\cite{singh_effect_2017} and hence creates an initial coherent superposition, which results in quick de-localization of excitation among the BChla sites of FMO complex. However, for an initial mixed state such an initial coherent superposition to enhance the de-localization of excitation energy is absent. Hence initial pure state relaxes more efficiently as compared to initial mixed sate. Our results here can be interpreted as reminiscent to the coherent control approach by Shapiro and Brumer, who have shown how to optimize the initial condition to obtain maximised output in photo-diisociation and several other photo-chemical processes\cite{Shapiro}.

\begin{figure}[!h]
\begin{center}
\subfloat{\label{Sink_277_3_Markovian}
\includegraphics[width = 3.0 in]{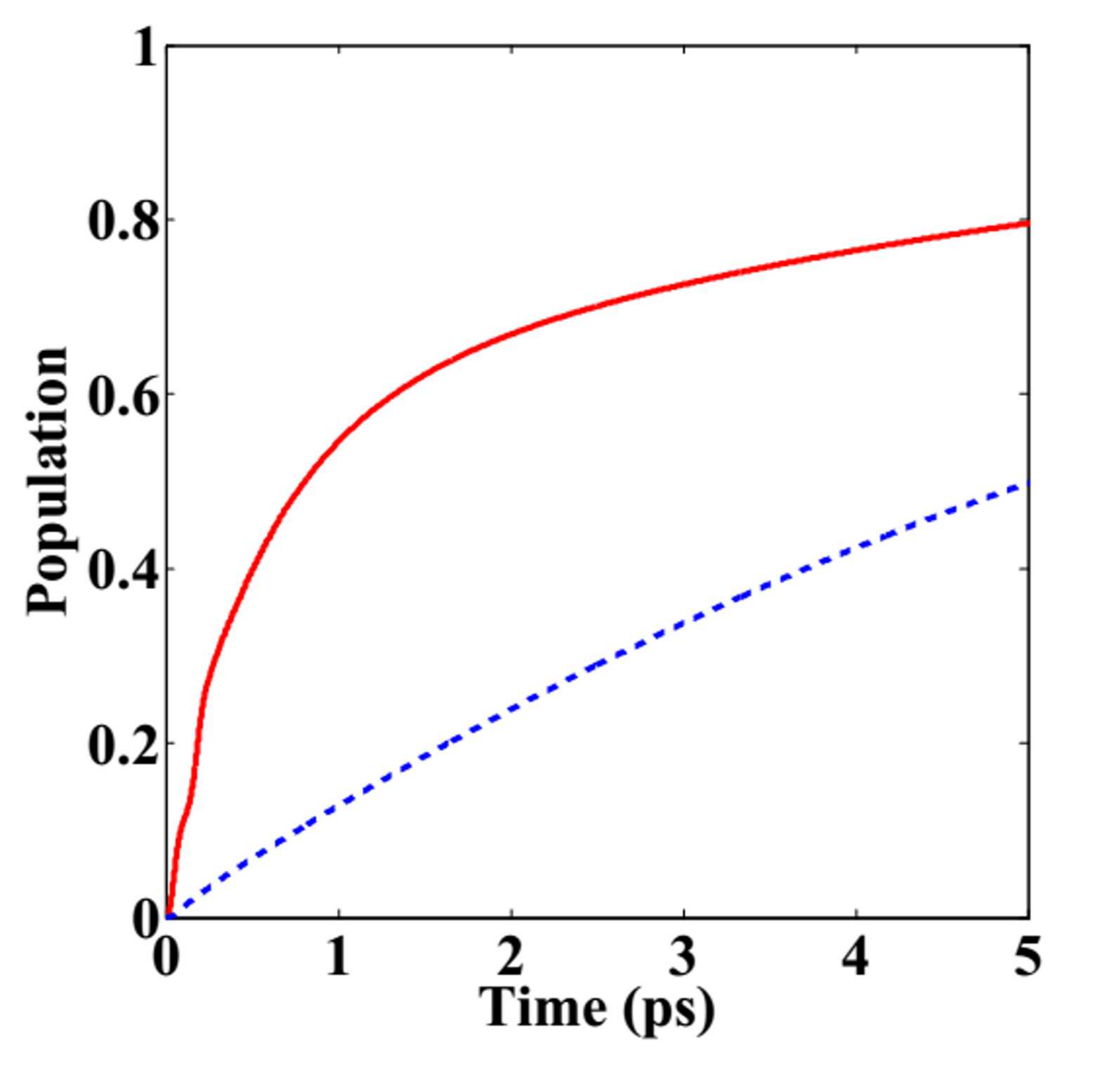}}
\caption{(Color online) Time evolution of population of sink with Markovian master equation. Here the solid (dashed) line shows the dynamics for initial pure state (initial mixed state). The de-phasing rate ($\gamma$) has been chosen to be 26 cm$^{-1}$.}
\label{Coherence Control}
\end{center}
\end{figure}

\subsection{Effect of inhomogeneous environment}
\begin{figure}[!h]
\begin{center}
\subfloat[]{\label{Sink_277_3_no_environment}
\includegraphics[width = 3.2 in]{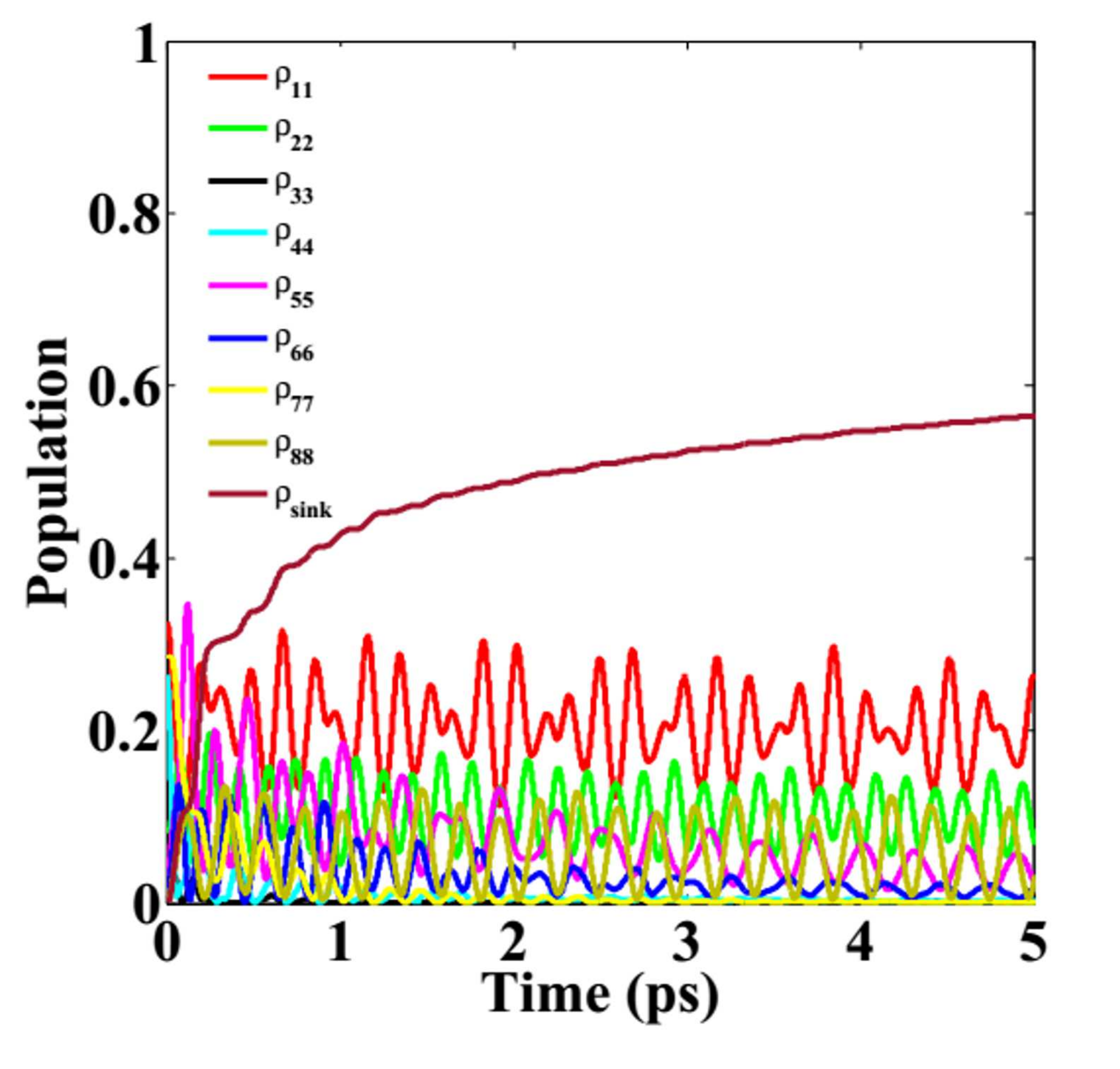}}
\quad
\subfloat[]{\label{Sink_277_3_strong_environment}
\includegraphics[width = 3.0 in]{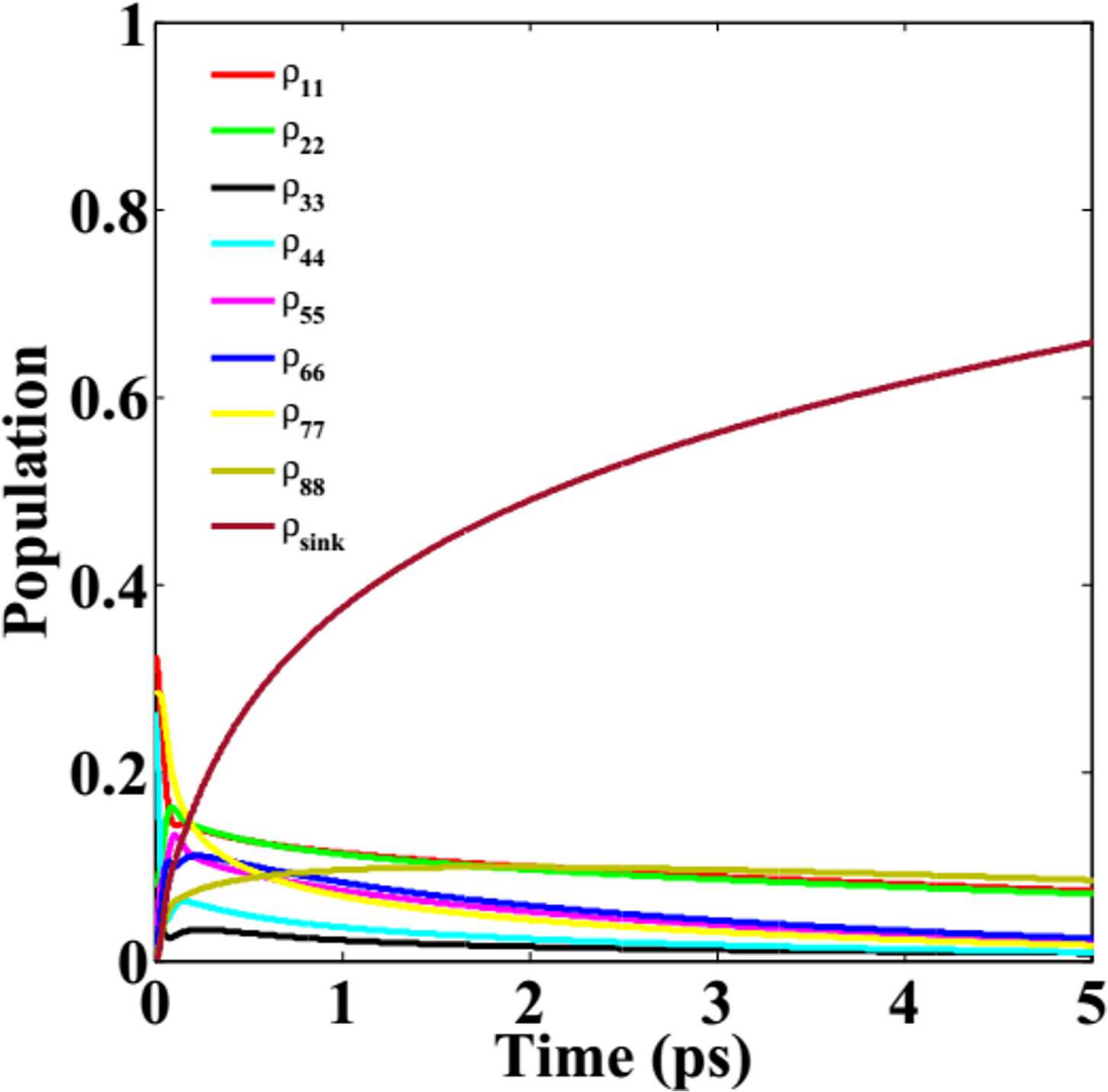}}
\caption{(Color online) Evolution of the excitation transfer dynamics at T = 277 K (a) without protein environment and (b) with strong protein environment. Here for the strong protein environment the electron-phonon coupling has been chosen to be five time stronger than the values given in Table\ref{Table_EET}.}
\label{Sink_277_3_environment}
\end{center}
\end{figure}

Next, to study the effect of protein environment on the coherently initiated ETE, we have displayed the dynamics in the absence of environment and in the presence of strong protein environment in Fig.\ref{Sink_277_3_environment}. Interestingly in both the cases, the ETE is less than in the case of the moderate coupling of the environment specified by Table\ref{Table_EET}. It indicates that to maximize the ETE, moderate coupling of environment is necessary as has already been observed as environmental assisted quantum transport (ENQT) process \cite{rebentrost_environment-assisted_2009,mohseni_environment-assisted_2008,rebentrost_role_2009}. From Fig.\ref{Sink_277_3_no_environment} it is clear that in the absence of environment, excitation oscillates even at the steady state. Hence in the absence of environmental decoherence, the dynamical coherence persists even at steady state. Presence of dynamical coherence at steady state creates the strong coherent superposition, which results in the re-distribution of excitation among all the BChla sites. It blocks the excitation transfer to RCC, hence minimizing the trapping at RCC. On the other hand, in the presence of strong protein environment the dynamical coherence disappears faster as shown in Fig.\ref{Sink_277_3_strong_environment}. Hence the dynamical coherent superposition exists for small time scale which led to the slow relaxation of excitation transfer due to stationary coherent superposition. However this slow relaxation of excitation also minimizes the transfer of excitation to RCC and hence the ETE as compared to the moderate environmental coupling. 

\section{Conclusion}

In conclusion, we have used realistic protein environment and several dominant vibronic modes with the realistic initial conditions to analyze the ETE. We observe that initial pure state enhances the ETE as compared to the initial mixed state. Moreover, we observe that in excitation transfer process the coherent superposition is required for an optimum time scale to get the maximum ETE. Hence optimum dephasing from protein environment is necessary for the efficient excitation transfer.

We further emphasize that to study the ETE explicitly, the RCC pigments must be included in the numerical simulation. We demonstrate that the interference between two channels 3$\rightarrow$9 and 4$\rightarrow$9 blocks the back transfer of excitation from RCC to FMO and hence helps in trapping the excitation efficiently. Although in an {\it isolated} FMO complex, the downward transfer of excitation exists and excitation get trapped at lowest energy site (i.e BChla 3), but in FMO-RCC complex this funnel model of excitation trapping seems to be incomplete without quantum interference.


\begin{acknowledgements}

This work was supported by Department of Science and Technology (DST), Govt. of India, under the grant number SR/S2/LOP-0021/2012.
\end{acknowledgements}


\nocite{*}
\bibliography{Paper_sink}     

\end{document}